\documentclass[a4paper,fleqn]{cas-dc}
\usepackage[numbers,square,sort&compress]{natbib}
\usepackage{tabularx}
\usepackage{placeins}
\usepackage{amsmath,amssymb,booktabs,array}

\makeatletter
\AddToHook{shipout/lastpage}{%
  \immediate\write\@auxout{\string\csxdef{lastpage}{\thepage}}}
\makeatother

\newcommand{\tabeq}[2]{%
  \begin{minipage}[c]{\linewidth}
  \setlength{\mathindent}{0pt}
  \setlength{\abovedisplayskip}{2pt}
  \setlength{\belowdisplayskip}{2pt}
  \begin{equation}#1\label{#2}\end{equation}
  \end{minipage}}

\ExplSyntaxOn
\RenewDocumentCommand \emailauthor { m m }
  {
    \int_gincr:N \g_ead_int
    \seq_gput_right:Nn \g_stm_ead_seq { { \ttfamily \tl_to_str:n { #1 } } }
  }
\RenewDocumentCommand \orcidauthor { m m }
  { \seq_gput_right:Nn \g_stm_orcid_seq { { \ttfamily \tl_to_str:n { #1 } } } }
\ExplSyntaxOff

\begin{document}
\let\WriteBookmarks\relax
\shorttitle{QA-based nesting in additive manufacturing}
\shortauthors{Wu et al.}
\title[mode=title]{Nesting in additive manufacturing using quantum annealing}
\author[1]{Xiangqian Wu}[orcid=0000-0003-4580-0813]
\cormark[1]
\ead{xiangqian.wu@rptu.de}
\author[1]{Jannik Gayer}
\author[1]{Marcel Wagner}
\author[1]{Jan C. Aurich}
\cortext[cor1]{Corresponding author}
\affiliation[1]{organization={Institute for Manufacturing Technology and Production Systems, RPTU Kaiserslautern-Landau},
  city={Kaiserslautern}, country={Germany}}

\begin{abstract}
Nesting in powder bed fusion (PBF), an additive manufacturing (AM) process, determines the positions and orientations of parts within the build space and influences build height and build time. As the number of parts and feasible placements increases, the nesting problem becomes computationally demanding. Exact approaches are limited to small problem sizes as computational effort increases rapidly with the problem size, while heuristic and metaheuristic approaches involve a trade-off between solution quality and computation time. Quantum annealing (QA) offers an alternative for solving combinatorial optimization problems, but its application to nesting in AM remains largely unexplored. This paper presents a QA-based approach for nesting in AM to investigate its feasibility and its scaling behavior across three problem sizes. An iterative binary quadratic model-based approach (IBQM) is developed to minimize build height while considering the geometric characteristics of AM parts. Geometric abstraction, iterative decomposition, and variable pruning are applied to address the computational complexity of the nesting problem. The approach is evaluated using three cases with increasing problem sizes and compared to a classical approach combining tabu search and simulated annealing (TBSA). Compared with TBSA, IBQM reduces the build height of the original parts by up to 13.32\% and the computation time by up to 74.27\%.
\end{abstract}

\maketitle

\section{Introduction}

The increasing industrial applicability of additive manufacturing processes establishes additive manufacturing (AM) as a production process for complex, customized parts \cite{ALAMI2023102516}. The growing variety and quantity of additively manufactured parts create new requirements for production planning and control (PPC), which must consider the specific characteristics of AM \cite{OH}. A central task of production control is nesting, since a build job comprises several parts manufactured together \cite{EmmelmannRudolph}.

In powder bed fusion (PBF), nesting determines the orientation and position of parts within the build space. All parts must remain within the build space without intersecting each other. The main objective is minimizing the build height, which determines the number of layers and thus the recoating time, the required powder volume and the build space utilization, whereas the exposure time of a given set of parts is largely independent of the arrangement \cite{Gaillard2024}. With an increasing number of parts and of feasible positions and orientations, the search space grows multiplicatively, resulting in an NP-hard combinatorial optimization problem \cite{Pinto}. In industrial practice, nesting is repeated for each build job with a different combination of parts, so that high solution quality must be achieved within acceptable computation times.

Exact, heuristic and metaheuristic approaches are applied to solve the nesting problem. Lamas-Fernandez et al. develop an exact voxel-based formulation of the three-dimensional packing problem \cite{Lamas-Fernandez}. Araújo et al. apply deepest-bottom-left-fill placement heuristics to arrange arbitrary 3D geometries \cite{Araujo}. Canellidis et al. optimize build space utilization by combining a genetic algorithm with a heuristic placement strategy \cite{CANELLIDIS}. Exact approaches provide reliable reference solutions but are limited to small problem sizes due to the strongly increasing computational effort. Heuristics and metaheuristics achieve shorter computation times but depend strongly on parameterization and the initial solution and are subject to a trade-off between solution quality and computation time. This limits their applicability to larger build jobs.

Quantum annealing (QA) is a metaheuristic within quantum computing designed to find near-optimal solutions in very short computation times \cite{McGeoch}. QA thus represents a promising approach to address these limitations \cite{Cohen}. The practical application of QA is facilitated by providers such as D-Wave\footnote{Naming of specific companies is done solely for the sake of completeness and does not necessarily imply an endorsement of the named companies or that their products are necessarily the most suitable for the intended purpose.}, which offers cloud access to its quantum annealers and enables the implementation and evaluation of QA-based optimization approaches on physical hardware \cite{dwaveLeapHybridDocs}. Since physical qubit count and connectivity limit directly solvable problem sizes, quantum-classical hybrid solvers combine classical and quantum resources to address this limitation \cite{dwave2023HybridQuadratic}. QA has already shown effectiveness for optimization problems of PPC such as layout planning \cite{WU2025627}, scheduling \cite{SCHWORM2024142}, and material flow planning \cite{GAYER202660}. In the production control of conventional manufacturing systems, QA has additionally been applied to a 3D bin-packing problem \cite{Rahmani}, in which spatial placement decisions are optimized without AM-specific constraints.

The use of QA for nesting in AM has hardly been investigated. Hence, a QA-based approach is proposed, primarily intended to demonstrate the feasibility of a QA-compatible formulation and its solution.

\begin{figure*}[!tp]
\centering
\includegraphics[width=\textwidth]{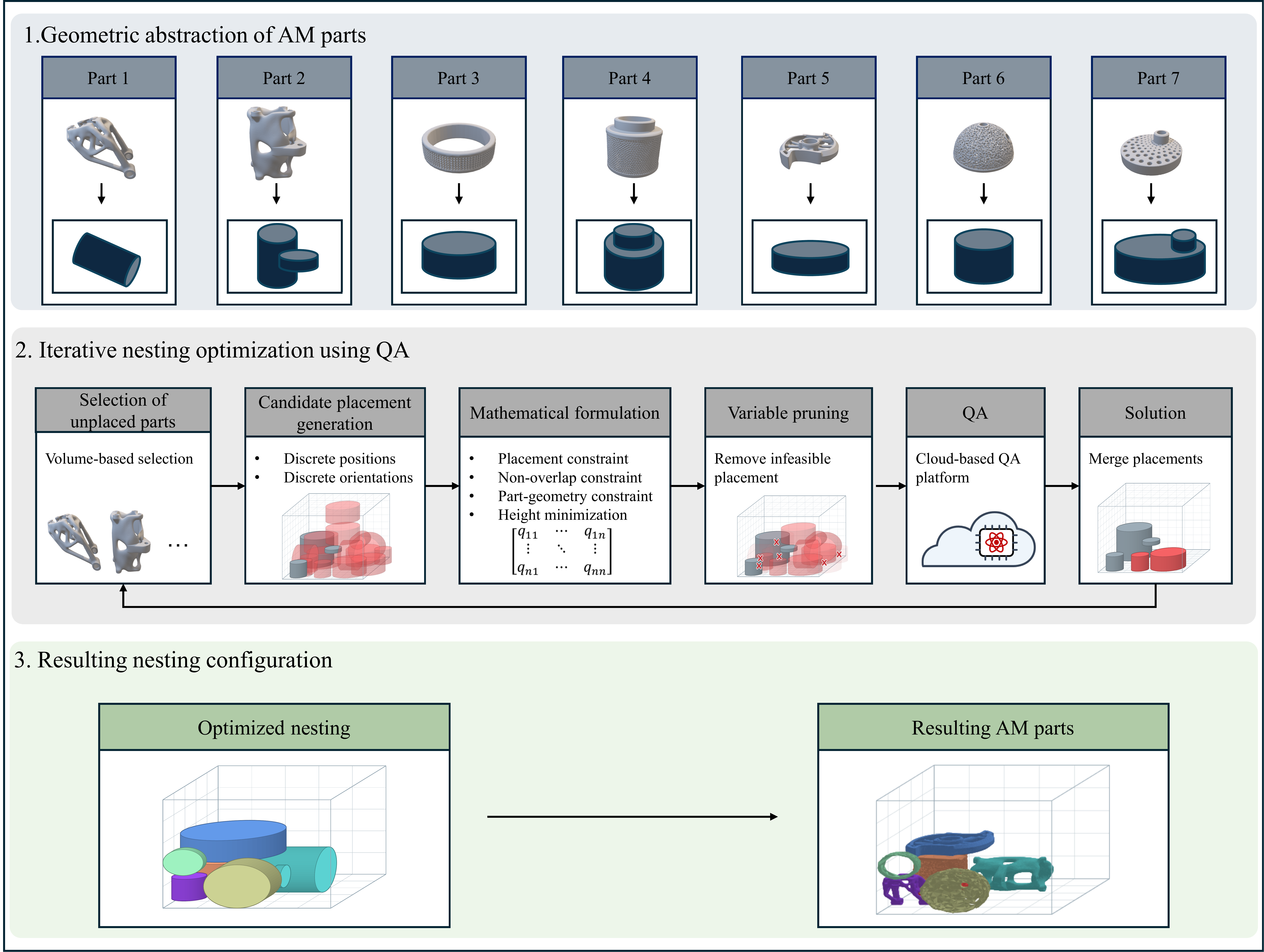}
\caption{Framework of the QA-based nesting approach IBQM.}
\label{fig:framework}
\end{figure*}

\section{Quantum-annealing-based nesting in additive manufacturing}
\subsection{Framework}
The proposed QA-based approach aims to determine the arrangement of AM parts within a predefined build space. As shown in Fig.~\ref{fig:framework}, the part geometries are represented by cylindrical approximations to reduce the geometric complexity. Depending on the part geometry, one or two cylindrical elements are used.

After geometric abstraction, iterative decomposition is applied to address the limitations of current QA hardware. In each iteration, a subset of unplaced parts is selected considering the part volume and feasible placements. Candidate positions and orientations are generated on the discretized build space. The available region is enlarged within the build-space limits only when necessary for feasibility. The placement, non-overlap and part-geometry constraints and the objective of reducing the build height define the nesting problem. These requirements are converted into a QA-based formulation.

Variable pruning removes infeasible placements to reduce the number of decision variables. The QA-based formulation is solved using a QA hybrid solver combining simulated annealing, tabu search, and QA. The obtained placements are fixed for the next iteration. This procedure continues until all parts are positioned. The cylinder positions and orientations are transferred to the original AM geometries. In this paper, the QA-based approach uses a binary quadratic model (BQM) formulation with an iterative approach and is denoted as IBQM.

\FloatBarrier
\subsection{Use case and formulation}
\label{sec:use_case}
Seven different AM part types, shown in Fig.~\ref{fig:framework}, are used to evaluate the proposed approach. Three cases with increasing problem size are investigated. Case~1 contains seven AM parts with one instance of each part type. Case~2 and Case~3 contain two and three instances of each part type, resulting in 14 and 21 AM parts, respectively. The seven part types are represented by ten cylindrical elements. The build space is $100\times100\times100$~mm and is discretized using a 10~mm grid $\mathcal G$. Candidate placements are considered for three orientations along the coordinate axes. Part geometries and cylindrical approximations can be found in the \hyperref[app:data]{supplementary material}.

\begin{table*}[pos=!tp]
\caption{BQM formulation.}
\label{tab:bqm_formulation}
\centering
\footnotesize
\renewcommand{\arraystretch}{1.1}
\renewcommand{\tabularxcolumn}[1]{m{#1}}
\begin{tabularx}{\textwidth}{@{}>{\raggedright\arraybackslash}p{2.2cm}>{\raggedright\arraybackslash}m{4.5cm}>{\raggedright\arraybackslash}X@{}}
\toprule
\textbf{Parameters} & AM parts & $a\in\mathcal P=\mathcal P_1\cup\mathcal P_2$\\
& Cylindrical elements of each AM part &
$\mathcal E_a=\begin{cases}
\{i\},&a\in\mathcal P_1\\
\{i,j\},&a\in\mathcal P_2,\ i<j
\end{cases}$\\
& Parts and elements selected in iteration $k$ &
$\mathcal P^{(k)}\subseteq\mathcal P,\quad
\mathcal I^{(k)}=\bigcup_{a\in\mathcal P^{(k)}}\mathcal E_a$\\
& Radius and length of element $i$ & $r_i,\ l_i$\\
& Distances from the center along $x,y,z$ &
$\displaystyle\boldsymbol d_{io}=\begin{cases}
(r_i,r_i,l_i/2),&o=0\\
(r_i,l_i/2,r_i),&o=1\\
(l_i/2,r_i,r_i),&o=2
\end{cases}$\\
& Feasible positions and orientations &
$p=(x_p,y_p,z_p,o_p)\in\mathcal F_i^{(k)}\subseteq\mathcal G\times\{0,1,2\}$\\
& Overlapping conditions of two elements &
$\displaystyle\mathcal C^{(k)}\subseteq\left\{(i,p,j,q)\ \middle|\
\begin{aligned}
&i<j,\ i,j\in\mathcal I^{(k)}\\
&p\in\mathcal F_i^{(k)},\ q\in\mathcal F_j^{(k)}
\end{aligned}\right\}$\\
& Relative displacements of elements of the same part &
$\boldsymbol\delta\in\Delta_{ij}^{o}\subset\mathbb R^3\times\{0\}$\\
\midrule
\textbf{Decision variables} & Position and orientation in iteration $k$ &
\tabeq{X_{ip}^{(k)}=\begin{cases}
1,&\text{if element }i\text{ is placed at }p\\
0,&\text{otherwise}
\end{cases}}{eq:variable}\\
\midrule
\textbf{Constraints} & Unique-placement constraint &
\tabeq{H_1^{(k)}=\sum_{i\in\mathcal I^{(k)}}\Bigl(1-\sum_{p\in\mathcal F_i^{(k)}}X_{ip}^{(k)}\Bigr)^2}{eq:unique}\\
& Non-overlap constraint &
\tabeq{H_2^{(k)}=\sum_{(i,p,j,q)\in\mathcal C^{(k)}}X_{ip}^{(k)}X_{jq}^{(k)}}{eq:overlap}\\
& Part-geometry constraint &
\tabeq{H_3^{(k)}=\sum_{\substack{a\in\mathcal P^{(k)}\cap\mathcal P_2\\\mathcal E_a=\{i,j\}}}
\sum_{p\in\mathcal F_i^{(k)}}X_{ip}^{(k)}
\Bigl(1-\sum_{\substack{\boldsymbol\delta\in\Delta_{ij}^{o_p}\\p+\boldsymbol\delta\in\mathcal F_j^{(k)}}}
X_{j,p+\boldsymbol\delta}^{(k)}\Bigr)}{eq:part_geometry}\\
\midrule
\textbf{Objective} & Minimization of the build height &
\tabeq{H_4^{(k)}=\sum_{i\in\mathcal I^{(k)}}\sum_{p\in\mathcal F_i^{(k)}}
(z_p+d^z_{io_p})X_{ip}^{(k)}}{eq:objective}\\
\midrule
\textbf{BQM} & &
\tabeq{H^{(k)}=
\lambda_1\widehat H_1^{(k)}+\lambda_2\widehat H_2^{(k)}+
\lambda_3\widehat H_3^{(k)}+w\widehat H_4^{(k)}}{eq:bqm}
\par\smallskip
with\par\nobreak\smallskip
\tabeq{\widehat H_m^{(k)}=
\frac{H_m^{(k)}}{\max\left\{1,\max_{\boldsymbol X^{(k)}}H_m^{(k)}\right\}},
\quad m\in\{1,2,3,4\}}{eq:normalization}\\
\bottomrule
\end{tabularx}
\end{table*}

The nesting problem is formulated as a BQM for iteration $k$, as summarized in Table~\ref{tab:bqm_formulation}.

Placements exceeding the build-space boundaries or intersecting an already positioned element are removed before the BQM variables in Eq.~\eqref{eq:variable} are generated. Hence, the binary variables are exclusively defined for feasible candidate placements.

Each cylindrical element must be assigned to exactly one position and orientation, which is ensured by the \textit{unique-placement constraint} in Eq.~\eqref{eq:unique}. Different AM parts must not intersect, which is ensured by the \textit{non-overlap constraint} in Eq.~\eqref{eq:overlap}. For parts represented by two cylindrical elements, the predefined relative arrangement must be maintained, which is ensured by the \textit{part-geometry constraint} in Eq.~\eqref{eq:part_geometry}.

To minimize the build height, the upper extent of each cylindrical element in the build direction is penalized in Eq.~\eqref{eq:objective}. The constraint and objective terms are normalized to $[0,1]$ using Eq.~\eqref{eq:normalization} and combined in the BQM with weights $\lambda_1$, $\lambda_2$, $\lambda_3$ and $w$ in Eq.~\eqref{eq:bqm}. All weights are set to one. Since the normalization in Eq.~\eqref{eq:normalization} renders the terms comparable, no weighting preference is imposed.

To explore the potential of QA, a classical hybrid method combining tabu search and simulated annealing (SA), denoted as TBSA, is applied. TBSA uses the same BQM formulation, iterative approach and classical search structure as IBQM. Since the aim is to demonstrate feasibility, each case is solved once with each approach, using one hybrid solver iteration per BQM. All cases use 25 SA reads with 10,000 sweeps per read and 20 tabu reads, with 300 reads on the quantum processing unit (QPU) for IBQM. Both approaches perform classical computations on an Intel Core i7-1165G7 processor with 16~GB RAM. The quantum resource for IBQM is the D-Wave Advantage2 system.

\FloatBarrier
\subsection{Results}
\label{sec:results}
\begin{figure*}[p]
\centering
\includegraphics[width=\textwidth]{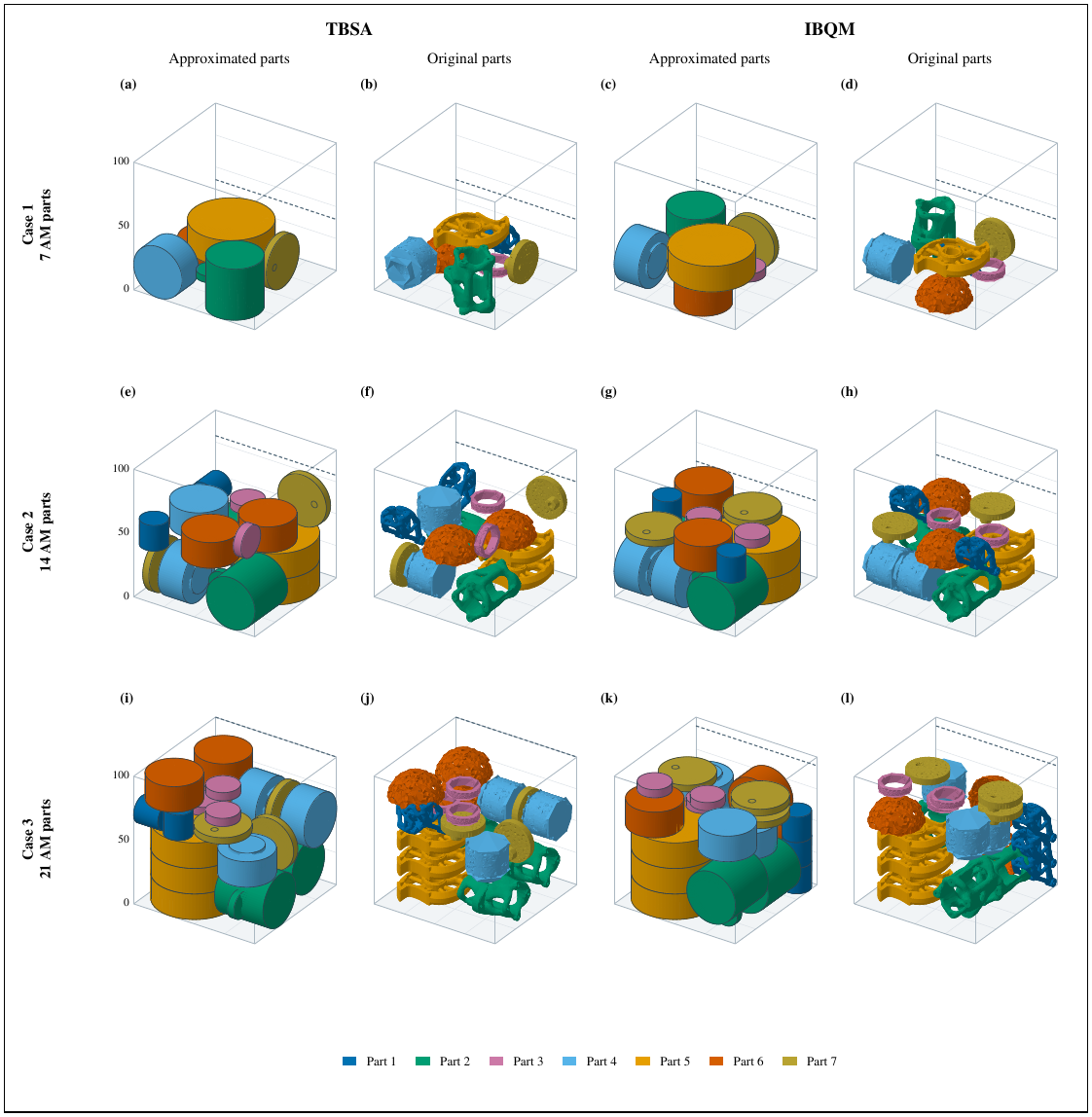}
\caption{Nesting configurations.}
\label{fig:results}
\end{figure*}

The nesting configurations obtained for the three cases are shown in Fig.~\ref{fig:results}, and the corresponding quantitative results are summarized in Table~\ref{tab:results}. Bold values indicate the lowest build heights and computation times in each case.

\makeatletter\global\@dbltopnum\z@\makeatother

\begin{table*}[pos=p]
\caption{Nesting results.}
\label{tab:results}
\centering
\footnotesize
\setlength{\tabcolsep}{3pt}
\begin{tabular*}{\textwidth}{@{\extracolsep{\fill}}lrrlrrrr@{}}
\toprule
Case & \shortstack{Number of\\variables} & \shortstack{Number of\\quadratic terms} & Approach & \shortstack{Build height of\\approximated parts\\in mm} & \shortstack{Build height of\\original parts\\in mm} & \shortstack{Computation time\\in s} & \shortstack{QPU access time\\in s}\\
\midrule
\multirow{2}{*}{Case 1} & \multirow{2}{*}{$14\,880$} & \multirow{2}{*}{$22\,397\,763$} & TBSA & \textbf{40.00} & \textbf{40.00} & 74.32 & --\\
 &  &  & IBQM & \textbf{40.00} & \textbf{40.00} & \textbf{19.12} & 0.165\\
\midrule
\multirow{2}{*}{Case 2} & \multirow{2}{*}{$29\,760$} & \multirow{2}{*}{$68\,412\,582$} & TBSA & 80.00 & 74.99 & 152.17 & --\\
 &  &  & IBQM & \textbf{60.00} & \textbf{65.00} & \textbf{43.45} & 0.578\\
\midrule
\multirow{2}{*}{Case 3} & \multirow{2}{*}{$44\,640$} & \multirow{2}{*}{$138\,044\,457$} & TBSA & 100.00 & 100.00 & 312.27 & --\\
 &  &  & IBQM & \textbf{93.15} & \textbf{93.10} & \textbf{117.80} & 1.148\\
\bottomrule
\end{tabular*}
\end{table*}

As the number of AM parts increases from seven to 21, the number of variables triples and the number of quadratic terms increases more than sixfold. The iterative structure of IBQM divides these large formulations into smaller BQMs, enabling the QA-based approach to solve all three cases.

Compared with TBSA, IBQM achieves the same build height in Case~1. For Case~2 and Case~3, the build heights of the original parts are reduced by 13.32\% and 6.90\%, respectively. The differences between the build heights of the original parts and their cylindrical approximations result from geometric abstraction.

The QPU access time refers to the runtime on quantum hardware and remains within seconds. Due to the classical structure, model submission and cloud transfer to the quantum annealer, the computation time of IBQM exceeds one minute in Case~3. However, TBSA takes approximately 2.65 times as long in this case. TBSA uses the same formulation and classical structure but requires more computation time in all cases and provides lower solution quality in the two larger cases, indicating a computational advantage of the QA-based hybrid approach for the investigated cases.

\FloatBarrier
\section{Conclusion}
\label{sec:conclusion}
This paper demonstrates the feasibility of the QA-based approach for nesting in AM. The results show that IBQM can solve increasing problem sizes while achieving equal or better solution quality and shorter computation times than the classical reference approach. However, each case was solved once, so that solution variance was not quantified, and scalability beyond the investigated problem sizes remains to be established.

Further research should evaluate larger nesting problems with more diverse part geometries, quantify solution variance across repeated runs, and compare the approach with established nesting methods. Dynamic scenarios, in which disturbances such as machine breakdowns require short-term rescheduling, remain to be addressed. Moreover, the integration of nesting and scheduling offers a direction for extending QA-based optimization to production control in AM.
\appendix
\section{Supplementary material}
\label{app:data}
The original part data and their cylindrical approximations associated with this article can be found online at 
https://seafile.rlp.net/d/40ab4fb203134241aed3/.

\section*{Acknowledgements}
This research was funded by the Ministerium f\"{u}r Wirtschaft, Tourismus, Energie und Klima
Rheinland-Pfalz within the project ``Quantflow'' (8401 RLP) and by the German Research
Foundation (DFG) within the project ``QuALyPlan'' (544507126).

\section*{Declaration of generative AI and AI-assisted technologies in the manuscript preparation process}
Generative AI (\emph{Claude Opus 5.5, Anthropic}) was used solely for improving language, style, and readability. All content of this paper was produced by the authors. The authors reviewed and approved all revised text and take full responsibility for this publication.

\bibliographystyle{model1-num-names}
\bibliography{cas-refs}

\begin{thebibliography}{10}
\expandafter\ifx\csname url\endcsname\relax
  \def\url#1{\texttt{#1}}\fi
\expandafter\ifx\csname urlprefix\endcsname\relax\def\urlprefix{Available from:
  }\fi
\expandafter\ifx\csname href\endcsname\relax
  \def\href#1#2{#2} \def\path#1{#1}\fi

\bibitem{ALAMI2023102516}
A.~H. Alami, A.~{Ghani Olabi}, A.~Alashkar, S.~Alasad, H.~Aljaghoub, H.~Rezk,
  M.~A. Abdelkareem, Additive manufacturing in the aerospace and automotive
  industries: Recent trends and role in achieving sustainable development
  goals, Ain Shams Engineering Journal 2023;14(11):102516.
\newblock \href {https://doi.org/10.1016/j.asej.2023.102516}
  {\path{https://doi.org/10.1016/j.asej.2023.102516}}.

\bibitem{OH}
Y.~Oh, P.~Witherell, Y.~Lu, T.~Sprock, Nesting and scheduling problems for
  additive manufacturing: A taxonomy and review, Additive Manufacturing
  2020;36:101492.
\newblock \href {https://doi.org/10.1016/j.addma.2020.101492}
  {\path{https://doi.org/10.1016/j.addma.2020.101492}}.

\bibitem{EmmelmannRudolph}
C.~Emmelmann, J.-P. Rudolph, F.~Güntzer, S.~Zühlke, Automatisierte
  arbeitsplanung und fertigungssteuerung für die additive produktion,
  Zeitschrift für wirtschaftlichen Fabrikbetrieb 2018;113(6):398--401.
\newblock \href {https://doi.org/10.3139/104.111925}
  {\path{https://doi.org/10.3139/104.111925}}.

\bibitem{Gaillard2024}
A.~Gaillard, S.~R{\"o}{\ss}ler, J.~Helmholz, S.~Junglas, W.~Boos, Optimierung
  der produktionsplanung und -steuerung f{\"u}r metallische additive fertigung,
  Zeitschrift f{\"u}r wirtschaftlichen Fabrikbetrieb 2024;119(9):608--614.
\newblock \href {https://doi.org/10.1515/zwf-2024-1110}
  {\path{https://doi.org/10.1515/zwf-2024-1110}}.

\bibitem{Pinto}
M.~Pinto, C.~Silva, M.~Th{\"u}rer, S.~Moniz, Nesting and scheduling
  optimization of additive manufacturing systems: Mapping the territory,
  Comput. Oper. Res. 2024;165(C).
\newblock \href {https://doi.org/10.1016/j.cor.2024.106592}
  {\path{https://doi.org/10.1016/j.cor.2024.106592}}.

\bibitem{Lamas-Fernandez}
C.~Lamas-Fernandez, J.~A. Bennell, A.~Martinez-Sykora, Voxel-based solution
  approaches to the three-dimensional irregular packing problem, Oper. Res.
  2023;71(4):1298–1317.
\newblock \href {https://doi.org/10.1287/opre.2022.2260}
  {\path{https://doi.org/10.1287/opre.2022.2260}}.

\bibitem{Araujo}
L.~J. Araújo, E.~Özcan, J.~A. Atkin, M.~Baumers, Analysis of irregular
  three-dimensional packing problems in additive manufacturing: a new taxonomy
  and dataset, International Journal of Production Research
  2019;57(18):5920--5934.
\newblock \href {https://doi.org/10.1080/00207543.2018.1534016}
  {\path{https://doi.org/10.1080/00207543.2018.1534016}}.

\bibitem{CANELLIDIS}
V.~Canellidis, J.~Giannatsis, V.~Dedoussis, Efficient parts nesting schemes for
  improving stereolithography utilization, Computer-Aided Design
  2013;45(5):875--886.
\newblock \href {https://doi.org/10.1016/j.cad.2012.12.002}
  {\path{https://doi.org/10.1016/j.cad.2012.12.002}}.

\bibitem{McGeoch}
C.~McGeoch, Adiabatic quantum computation and quantum annealing: Theory and
  practice, Synthesis Lectures on Quantum Computing 2014;5:1--93.
\newblock \href {https://doi.org/10.2200/S00585ED1V01Y201407QMC008}
  {\path{https://doi.org/10.2200/S00585ED1V01Y201407QMC008}}.

\bibitem{Cohen}
E.~Cohen, B.~Tamir, D-wave and predecessors: From simulated to quantum
  annealing, International Journal of Quantum Information 2014;12:1430002.
\newblock \href {https://doi.org/10.1142/S0219749914300022}
  {\path{https://doi.org/10.1142/S0219749914300022}}.

\bibitem{dwaveLeapHybridDocs}
{D-Wave Systems Inc.},
  \href{https://docs.dwavequantum.com/en/latest/leap_sapi/index.html}{{Leap
  Service}}, \url{https://docs.dwavequantum.com/en/latest/leap_sapi/index.html}
  [accessed 28 October 2026].

\bibitem{dwave2023HybridQuadratic}
{D-Wave Systems Inc.},
  \href{https://www.dwavequantum.com/resources/white-paper/hybrid-solvers-for-quadratic-optimization/}{{Hybrid
  solvers for quadratic optimization}}, White paper,
  \url{https://www.dwavequantum.com/resources/white-paper/hybrid-solvers-for-quadratic-optimization/};
  2022 [accessed 28 October 2026].

\bibitem{WU2025627}
X.~Wu, P.~Schworm, M.~Klar, J.~C. Aurich, Factory layout planning using quantum
  annealing, CIRP Annals 2025;74(1):627--631.
\newblock \href {https://doi.org/10.1016/j.cirp.2025.03.013}
  {\path{https://doi.org/10.1016/j.cirp.2025.03.013}}.

\bibitem{SCHWORM2024142}
P.~Schworm, X.~Wu, M.~Klar, M.~Glatt, J.~C. Aurich, Multi-objective quantum
  annealing approach for solving flexible job shop scheduling in manufacturing,
  Journal of Manufacturing Systems 2024;72:142--153.
\newblock \href {https://doi.org/10.1016/j.jmsy.2023.11.015}
  {\path{https://doi.org/10.1016/j.jmsy.2023.11.015}}.

\bibitem{GAYER202660}
J.~Gayer, M.~Wagner, J.~C. Aurich, A modeling approach of material flow
  optimization in manufacturing using quantum annealing, Procedia CIRP
  2026;146:60--65, cIRP Conference on Manufacturing Systems (CIRP CMS).
\newblock \href {https://doi.org/10.1016/j.procir.2026.03.194}
  {\path{https://doi.org/10.1016/j.procir.2026.03.194}}.

\bibitem{Rahmani}
M.~Rahmani, N.~Umang, M.~Christensen, {Application of hybrid classical-quantum
  annealing technology to the 3D Bin-Packing Problem} 2024.
\newblock \href {http://arxiv.org/abs/2408.15365} {\path{arXiv:2408.15365}}.

\end{thebibliography}
\end{document}